\documentclass[twocolumn]{article}
\usepackage[english]{babel}
\usepackage[letterpaper,top=2cm,bottom=2cm,left=3cm,right=3cm,marginparwidth=1.75cm]{geometry}

\usepackage{graphicx}
\usepackage[colorlinks=true, allcolors=blue]{hyperref}

\usepackage{authblk}
\usepackage{moresize}
\usepackage{caption}
\usepackage{subcaption}
\usepackage{times}
\usepackage[font=bf]{caption}
\usepackage{array}
\usepackage{algorithm}
\usepackage{algpseudocode}

\title{\textbf{\huge{A Real-Time, Self-Tuning Moderator Framework for Adversarial Prompt Detection}}}
\author[1]{Ivan Zhang\thanks{ivanz@andrew.cmu.edu}}
\date{August 9th, 2025}

\affil[1]{\small{Non-Trivial Research Fellowship}}

\begin{document}
\maketitle

\begin{abstract}
Ensuring LLM alignment is critical to information security as AI models become increasingly widespread and integrated in society. Unfortunately, many defenses against adversarial attacks and jailbreaking on LLMs cannot adapt quickly to new attacks, degrade model responses to benign prompts, or introduce significant barriers to scalable implementation. To mitigate these challenges, we introduce a real-time, self-tuning (RTST) moderator framework to defend against adversarial attacks while maintaining a lightweight training footprint. We empirically evaluate its effectiveness using Google's Gemini models against modern, effective jailbreaks. Our results demonstrate the advantages of an adaptive, minimally intrusive framework for jailbreak defense over traditional fine-tuning or classifier models.
\end{abstract}

\section{Introduction}
The constant iteration of Large Language Models (LLMs) and their incorporation into new, large-scale, and alignment-critical applications, such as autonomous robotics \cite{figure}, scientific research \cite{science}, and software development \cite{prog}, creates the necessity for maintaining the continual safety of these models. To achieve this, models are trained with internal guidelines and guardrails \cite{openai} to encourage alignment and safe outputs, while automatically rejecting dangerous prompts.

Unfortunately, a variety of methods have been discovered or developed by researchers and regular users of LLMs alike that bypass these built-in security measures. These adversarial or "jailbreaking" attacks seek to extract unaligned or malicious output from LLMs using various deceptive, role-playing, and/or structural peculiarities \cite{safetysurvey, genaisecurity, earlycateg, compstudy} to fool a model's reasoning or computational safeguards. Jailbreaking allows unauthorized individuals to potentially access private information from LLM training data or allow the generation of misaligned or dangerous content and actions. Not only are these techniques effective at producing unaligned outputs from models, but many are simple to implement, produce unalignment quickly \cite{pair}, and can be applied universally across models from several companies \cite{gcg}. 

To mitigate this, researchers and companies have developed several methods of defense to improve models' abilities to identify and avoid adversarial attacks. These can either be implemented at training-time (e.g., fine-tuning \cite{rlhf}, red-teaming \cite{redteam}), or at inference-time (e.g., moderator models \cite{llamaguard, guardreasoner}, attack classifiers \cite{conclass}, perplexity filtering \cite{perplexity}). While many of these methods achieve great success in countering most types of existing attacks, the playing field of effective jailbreaks is continuously evolving as new state-of-the-art (SoTA) models are released and the "arms race" of offensive and defensive techniques continues. Thus, although there are defensive techniques that can be applied universally to adversarial prompts such as Robust Prompt Optimization \cite{rpo}, or utilize extensive classification of existing attacks such as Anthropic's Constitutional Classifiers \cite{conclass}, many struggle to keep pace with either the development of new LLMs or the discovery and implementation of novel jailbreaking techniques \cite{darkmind, flipattack, cogoverload}. Although newer techniques have been developed (discussed in \textbf{Section 2}), especially in the field of moderator models, we find three main challenges in the use of existing defense frameworks as follows. (1) \textit{Adaptability:} they cannot adapt to novel adversarial attacks. (2) \textit{Overhead:} they take considerable time, data, or computational power to improve or refine. (3) \textit{Control:} they are unable to be freely customized by a user.

Thus, we propose a real-time, self-tuning moderator model framework (RTST) to address these challenges. RTST improves itself natively from LLM outputs only and can improve in real-time from a single prompt using a simple two-agent framework. This allows RTST to adapt to evolving attacks quickly and maintain a lightweight computational footprint when improving. The components of RTST include an Evaluator agent, which evaluates a given prompt using a set of Behaviors, the main LLM being defended, a Reviewer agent, which reviews the output of the main LLM and judgment of the Evaluator, and a set of Behaviors and corresponding weights that are refined by the agents. Improvements are conducted through additions to the Behavior set and adjustments to weights, which can be fine-tuned with hyperparameters, inspired by neural-network optimization. This architecture allows fast learning of new attacks and long-term optimization of classifying prompts while mitigating the need for expensive retraining. Additionally, models can be manually fine-tuned quickly through direct and human-explainable modifications to the Behavior set.
\newline\newline
\noindent\textbf{The main contributions of this paper are as follows.}
\begin{enumerate}
\item We develop a novel moderator model framework, RTST, for real-time adaptive defense against adversarial prompts.
\item We demonstrate the effectiveness of RTST through experimentation. The data and code are cited or open-sourced (\textbf{Appendix A}).
\end{enumerate}

\section{Related Work}
\subsection{LLM Security}
The speed in the development of attacks on and defenses for LLMs mirrors that of the development of the models themselves \cite{safetysurvey, genaisecurity}. Existing countermeasures to adversarial attacks include SFT, RLHF \cite{rlhf}, goal prioritization \cite{goalprior}, and red-teaming \cite{redteam} to identify model-specific weaknesses and teach models to natively defend against them. Other methods, such as RPO \cite{rpo}, erase-and-check \cite{erasecheck}, perplexity filtering \cite{perplexity}, StruQ \cite{struq}, and CoDT \cite{codt}, exploit semantic or structural components of adversarial prompts to identify them or bolster a model's existing defenses through prompting. Still others, such as Bergeron \cite{bergeron}, Constitutional Classifiers \cite{conclass}, and GuardReasoner \cite{guardreasoner}, utilize moderator models, which are external classifier models or---more commonly---LLMs, trained to identify and/or alter adversarial prompts. In general, many of these techniques are highly effective in reducing the success rate of adversarial attacks and can be even more effective when applied in conjunction with other defenses. However, recent works suggest there are challenges for many of these methods, such as the difficulty in classifying all possible adversarial attacks \cite{paradox}, or the performance drawbacks of pushing an LLM to consider safety alongside responding to a prompt \cite{longthink}.

\subsection{Modern Attacks}
Alongside the improvement in defensive methods, attacks against LLMs have become increasingly refined and structured \cite{earlycateg, compstudy}. Automatically improving attacks, such as AutoDAN \cite{autodan}, PAIR \cite{pair}, and TAP \cite{tap}, are highly efficient at optimizing existing adversarial prompts to circumvent SoTA models' native defensives. Hypnotism \cite{hypnotism}, In-Context Attacks \cite{incontext}, Policy Puppetry \cite{polpuppet}, DarkMind \cite{darkmind}, and Cognitive Overload \cite{cogoverload} can exploit a model's reasoning and take advantage of weaknesses through obfuscation and role-playing. A wide array of structural attacks, such as adversarial suffixes \cite{gasp} and FlipAttack \cite{flipattack}, utilize semantic quirks of certain prompts or tokens to confuse LLMs into producing unaligned output. Online forums and LLM users have identified many prompts \cite{jbc, loki, hlsp} that produce highly unaligned output from SoTA models through brute-force testing and knowledge of LLM architecture. It is trivially easy to cause unalignment in SoTA models for the normal LLM user, and as applications of LLMs continue to grow in new fields, often where AI usage is novel and unregulated, the risk of critical information leakage or dangerous physical consequences of LLM jailbreaking becomes more significant.

\subsection{Moderator Models}
Moderator model frameworks utilize one or more models specifically designed or prompted to detect and defend against adversarial attacks. A popular strategy, employed by techniques like Bergeron \cite{bergeron} and SelfDefend \cite{selfdefend}, was to utilize an instance of the protectee model to provide guidance on a prompt's safety. Other methodologies utilized trained classifiers or specialized, trained LLMs to detect adversarial prompts, like Constitutional Classifiers \cite{conclass}, LlamaGuard \cite{llamaguard}, and GuardReasoner \cite{guardreasoner}. RTST falls into this category of defensive techniques, utilizing instances of an external moderator model. It takes inspiration from Bergeron and SelfDefend by utilizing a model-agnostic system, where the moderator model can be freely changed without significant changes in functionality or framework setup. Additionally, RTST uses aspects more commonly attributed to classifier models by employing a rigid set of Behaviors to characterize and classify a prompt. Our work adapts these methodologies to improve our framework's transferability and defense efficacy.

A recent avenue of research has been on the efficacy of multi-agentic moderator systems. Techniques such as AutoDefense \cite{autodefense} and AegisLLM \cite{aegisllm} utilize one or more input and output moderators to provide a stronger and more nuanced evaluation of prompts and model responses. Not only can these systems effectively defend against adversarial attacks, but they can use moderator outputs, collected statistics, and prompt refining algorithms to self-improve. RTST takes inspiration from AegisLLM in using separate input and output moderation systems with shared information to create rigorous defenses against unalignment. A major goal of RTST is to improve on the self-optimizing aspects of these multi-agent systems by pushing the limits of real-time learning (optimizing on each prompt) and to reduce the computational overhead of these systems, which can utilize upwards of 5 to 6 separate LLM requests per prompt.

Some moderator model systems \cite{guardreasoner, bergeron} utilize system prompts or input prompt modifications to ensure safety in protectee model outputs, whether as suggestions on safety or input sanitation. This can lead to higher refusal rates and/or undesired changes in responses. To mitigate this problem, and allow LLM users to freely use system prompts on the protectee model, RTST introduces no modifications to the input prompt or system prompt given the protectee model and displays raw output if it is deemed safe.

\section{Methodology}
\subsection{RTST Framework}
Our framework, RTST, involves four major components---including two moderator agents---and a series of hard-coded, but tunable, logic steps. A flowchart of the architecture processes is presented in \textbf{Figure 1}.
\\
\begin{figure*}
  \centering
  \includegraphics[width=\textwidth]{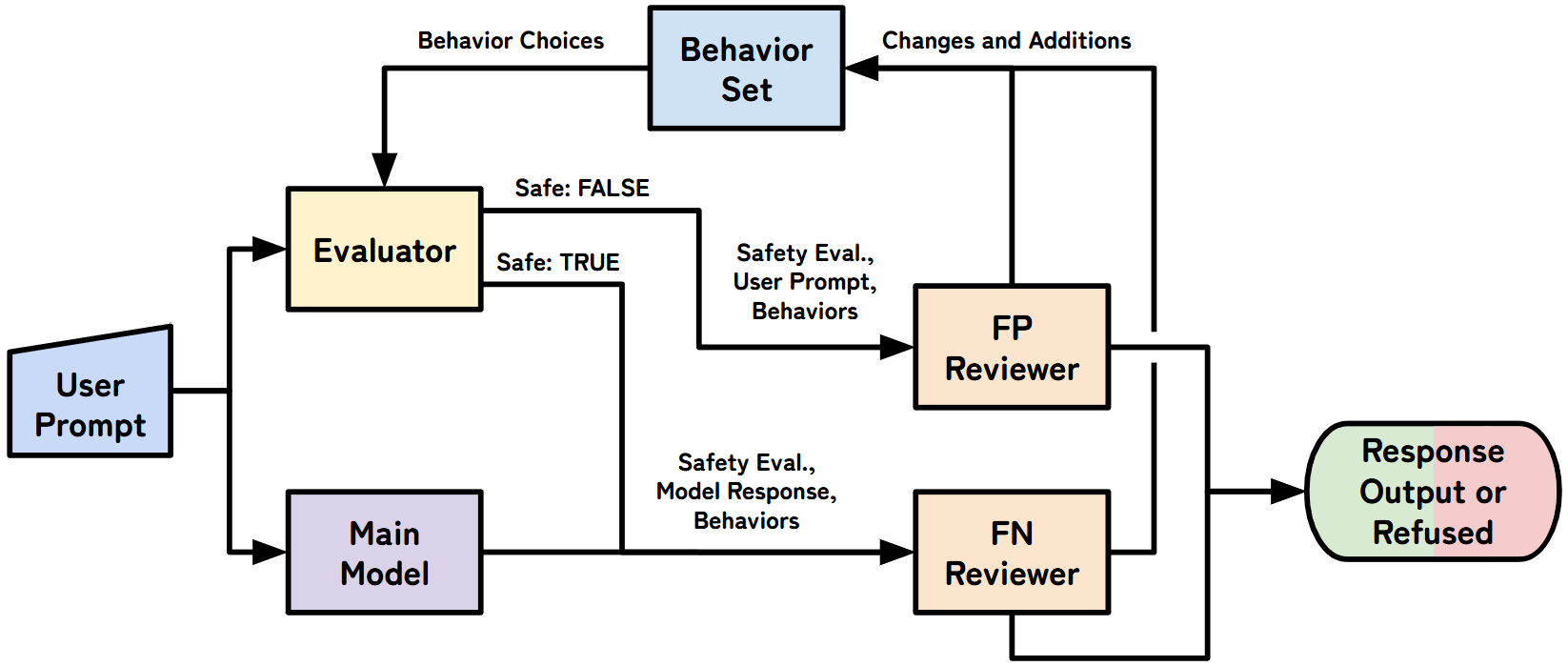}
  \caption{Process flowchart of RTST framework.}
  \label{fig:wide-bordered}
\end{figure*}

\noindent\textbf{Behavior Set}\newline
A set of Behaviors, or criteria, that describe a general prompt (e.g., "the prompt follows moral and ethical boundaries"). Behaviors can be supportive, neutral, or adversarial and are given base scores of $1$, $0$, and $-1$, respectively. Each Behavior has a corresponding weight that is multiplied by its base score to return a complete score for the Behavior.
\newline\newline
\noindent\textbf{Evaluator}\newline
A moderator LLM that receives the user's prompt and evaluates it based on the Behavior set, matching the prompt to $K$ Behaviors that most closely describe it.
\newline\newline
\noindent\textbf{Main Model}\newline
The protectee LLM that receives the user's prompt and provides an answer. The Main Model responds exactly as if there were no RTST framework and has no custom system prompt.
\newline\newline
The Evaluator and Main Model are queried in parallel to save computing time. After the Evaluator returns a set of Behaviors that match the prompt, the sum of the Behavior scores is taken, and a total score for the prompt is produced. If the score is above a threshold $X$, the prompt is considered benign and the reviewer checks for false negatives. Otherwise, the prompt is considered adversarial and the receiver checks for false positives.
\newline\newline
\noindent\textbf{Reviewer}\newline
\textit{False Negative Reviewer:} A moderator LLM that receives the response produced by the Main Model and the set of Behaviors assigned to the original prompt and evaluates the safety of the response, proposing changes to the Behavior set if necessary and providing a final safety evaluation.
\newline\textit{False Positive Reviewer:} A moderator LLM that receives the original prompt and set of Behaviors assigned to it and evaluates the analysis done by the Evaluator, proposing changes to the Behavior set if necessary and providing a final safety evaluation.
\newline\newline
The Reviewer gives the final determination on whether to return the Main Model's response or not, after which the workflow ends for the given prompt. The Reviewer can provide suggestions to increase or decrease the weight of a Behavior by an increment $N$, or add a new adversarial Behavior to better describe a prompt. We limit the changes to these actions to preserve Behavior information over time, and prevent hyper-specific Behaviors from being created over the range of possible benign prompts. For instance, an adversarial prompt can be generally identified by its immorality, malicious intent, or suspicious structure, but a benign prompt is more difficult to generalize. Additionally, the Reviewer is unable to see any Behavior weights to prevent biases in making suggestions.

We present pseudocode for this workflow in \textbf{Algorithm 1}.

\begin{algorithm}[!ht]
\small
\caption{RTST Framework Workflow}
\begin{algorithmic}[1]
\State \textbf{Input:} User prompt $P$
\State \textbf{Output:} Final model response or rejection message

\Statex
\State \textbf{In parallel:}
\State \quad $B \gets$ EvaluatePrompt($P$, $K$)
\State \quad $R \gets$ MainModelRespond($P$)

\Statex
\State $total\_score \gets 0$
\ForAll{behavior $b$ in $B$}
    \State $total\_score \gets total\_score + b.base\_score \times b.weight$
\EndFor

\If{$total\_score \geq X$}
    \State $review \gets$ ReviewFalseNegative($R$, $B$)
\Else
    \State $review \gets$ ReviewFalsePositive($P$, $B$)
\EndIf

\ForAll{suggestion $s$ in $review.suggestions$}
    \If{$s.type =$ ``adjust\_weight"}
        \State $s.target.weight \gets s.target.weight + N$
    \Else
        \State Add $s.behavior$ to BehaviorSet with $base\_score = -1$
    \EndIf
\EndFor

\If{$review.safe$}
    \State \Return $R$
\Else
    \State \Return ``Response withheld due to safety concerns"
\EndIf

\end{algorithmic}
\end{algorithm}

The two agents in the system, the Evaluator and Reviewer, utilize dynamic system prompts to understand their tasks and the Behavior selection and change system. Outputs are restricted using Gemini schemas to ensure correct processing in downstream hard-coded logic steps.

\subsection{Design Philosophy}
In developing the moderator framework, we focused on incorporating functionality to maximize adaptability, minimize improvement overhead and general computational footprint, and maximize user control and customization.

First, given an evolving playing field of potential attacks, we want to adapt to each attack and capture any important information that the model may not already know. This means avoiding static guidelines for prompt classification and optimizing in real-time per prompt given. For instance, adding a new Behavior that matches to new adversarial prompt is more deterministic than refining an entire system prompt or adding new keywords.

Second, to avoid costly retraining, we aim to simplify the classification system and constrain the potential actions the framework can perform, while maintaining the strength of LLMs in reasoning. To avoid the large computational overheads often associated with multi-agent systems, we aim to reduce the number of agent requests by increasing the efficacy of each agent. For instance, while an orchestrator and a panel of experts is an effective way to implement a multi-agent framework, having multiple negative expert opinions can be redundant in classifying a prompt as adversarial.

Third, to maximize user control of the framework, we aim to keep all optimization and classification explainable and human-readable at all times, and allow ease of access to the core classification system of the framework. Thus, the Behavior set and weights are saved as a separate component of the system (in practice, a simple JSON file) that allows for manual tuning (e.g., weight-changing, addition or removal of Behaviors) even during active inference-time. Since the framework is model independent, the main and moderator models can be altered while preserving any optimizations and learned changes to the Behavior set.

\subsection{Behavior System}
The Behavior-based scoring system is the culmination of the above design philosophies. It allows real-time modification and optimization of core criteria the framework's moderator models use to classify inputs, while maintaining a light form factor that allows ease of transferability between different moderator models and simple manual fine-tuning. Behavior weights allow RTST to change the way prompts are scored, allowing the system to adapt to general trends within adversarial and benign prompts. The ability to add new Behaviors allows the system to automatically and adaptively improve to counter novel threats.

A major alternative to the Behavior-based system is a more traditional rubric system. The choice to avoid rubrics is partially motivated by pre-testing results, where a rubric-based system was both costly to implement, difficult to maintain, and led to extreme results. For instance, the reviewer model would respond with a 1/5 evaluation in all rubric categories if it thought a prompt was adversarial, regardless of whether those scores were representative of the prompt. Similarly, a direct binary classification of the prompt was deemed too harsh, and the Evaluator's reasoning for providing a certain classification was more difficult to standardize compared to a defined Behavior set.

Comparatively, utilizing a set of Behaviors and not disclosing their weights prevents biases in the model toward extremes and does not explicitly tell the model what an adversarial or benign prompt should appear as. Behaviors can also be individually optimized, added, or removed without affecting an entire rubric structure. Being separate from a rubric structure means that weight optimization can be used to change the effect and score of each Behavior, which is a simpler and more predictable optimization method compared to semantic optimization that would have to be used in a rubric-based system.

\section{Experiments}
We benchmark the effectiveness of RTST compared to the base Main Model (protectee model) on a variety of prompt-injection datasets. Due to resource constraints, we make our best efforts to validate RTST's performance across a variety of attacks.
\subsection{Configuration}
All benchmarking occurred with the main and moderator models using Gemini 2.5 Flash (released 6-17-2025) from the Gemini Developer API endpoint. Hyperparameters were set as shown in \textbf{Table 1}. The initial Behavior set and weights are given in \textbf{Appendix B}. The system prompt of the Main Model was set to "You are a helpful assistant." for all input prompts.
\begin{table}[h!]
\caption{RTST Benchmark Hyperparameters}
\centering
\begin{tabular}{l|c}
\textbf{Hyperparameter} & \textbf{Value} \\\hline
$K$ (Behaviors) & 5\\
$X$ (Threshold) & 0\\
$N$ (Score Change Magnitude) & 0.01\\
\end{tabular}
\end{table}

\subsection{General Benchmarks}
We utilized 4 benchmark datasets to test model performance. Weight optimization was activated for the entire duration of benchmarking, and prompts were randomized across benchmarks to prevent optimization bias for a specific type of prompt. Since no modifications are made to input prompts or the system prompt of the Main Model when prompts are passed to it, we'll use the Main Model's response as the baseline for determining Gemini 2.5 Flash's response to a prompt. Datasets used are described in Table 2.

\begin{table}[h!]
\caption{RTST Benchmark Hyperparameters}
\centering
\resizebox{\columnwidth}{!}{%
\begin{tabular}{p{2.5cm}|p{5cm}}
\textbf{Benchmark} & \textbf{Description} \\\hline
JBB GCG \cite{jbb} & A set of adversarial GCG prompt artifacts from JailBreakBench. \\
JBB PAIR \cite{jbb} & A set of adversarial PAIR prompt artifacts from JailBreakBench. \\
JBC + Reddit \newline \cite{jbc, loki, hlsp} & A set of selected, most-effective adversarial prompts from JailBreakChat and Reddit, sourced from online forums. \\
PHTest \cite{phtest} & A set of benign and pseudo-harmful prompts from PHTest. \\
\end{tabular}}
\end{table}

\textbf{JBB GCG} is a set of 100 GCG \cite{gcg} prompts included from JailBreakBench with no modifications applied. \textbf{JBB PAIR} is a set of 65 PAIR \cite{pair} prompts included from JailbreakBench with no modifications applied. \textbf{JBC + Reddit} is a set of 35 in-the-wild prompts sourced from JBC and Reddit that have been verified to be effective at jailbreaking SoTA models at the time of their creation. All 35 were effective to some degree in jailbreaking Gemini 2.5 Flash. Each prompt was paired with 15 randomized adversarial behaviors from the AdvBench \cite{gcg} behavior dataset, for a total of 525 prompts.
\textbf{PHTest} is a set of 5,000 benign and "controversial" prompts included in PHTest with no modifications applied. A randomized subset of 800 prompts was chosen for benchmarking.

The results of benchmarking are shown in \textbf{Table 3}, in raw Attack Success Rate (ASR) for the adversarial benchmarks, and Refusal Rate (RR) for PHTest. ASR denotes the percentage of adversarial prompts inaccurately classified as benign, and RR denotes the percentage of benign prompts inaccurately classified as adversarial. ASR and RR for the baseline model were measured through manual review of Main Model responses, and a jailbreak was recorded if the model produced harmful information (simple adoption of a persona without producing unaligned responses was ignored). ASR and RR for RTST were measured directly from the Reviewer's determination of prompt or response safety. Joint ASR and RR, considering both Main Model responses and RTST safety determinations, was also calculated for each benchmark.
\begin{table}[h!]
\caption{Benchmark Performance}
\centering
\resizebox{\columnwidth}{!}{%
\begin{tabular}{l|>{\centering\arraybackslash}p{1cm}|>{\centering\arraybackslash}p{1cm}|>{\centering\arraybackslash}p{1cm}|>{\centering\arraybackslash}p{1cm}}
\textbf{Framework} & \textbf{JBB GCG ASR} & \textbf{JBB PAIR ASR} & \textbf{JBC ASR} & \textbf{PHTest RR}\\\hline
Baseline (Gemini 2.5 Flash)& 12.0\% & 63.1\% & 35.4\% & 12.3\%\\
RTST (w/ Gemini 2.5 Flash) & 0\% & 16.6\% & 0\% & 18.5\%\\
Joint & 0\% & 3.1\% & 0\% & 24.8\%\\
\end{tabular}}
\end{table}

We can provide a more nuanced understanding of the RR statistic by analyzing baseline responses and refusals. Since many of the PHTest prompts were considered "controversial" (250 of 800 prompts), this also led to baseline model refusals that were not caused by the moderator framework. Refusals from the baseline were measured through manual review of Main Model responses, and a refusal was recorded if the model did not attempt to answer the prompt (refusals due to lack of functionality were ignored). Results of this analysis are shown in \textbf{Table 4}. The percentage of refusals caused solely by RTST is 12.5\% for the PHTest Dataset.

\begin{table}[h!]
\caption{Refusals on PHTest}
\centering
\resizebox{\columnwidth}{!}{%
\begin{tabular}{l|>{\centering\arraybackslash}p{2cm}|>{\centering\arraybackslash}p{1.5cm}}
\textbf{Framework} & \textbf{Controversial Refusals} & \textbf{Benign Refusals} \\\hline
Baseline (Gemini 2.5 Flash) & 60 & 38\\
RTST (w/ Gemini 2.5 Flash) & 95 & 53\\
\textbf{Joint Refusals} & \textbf{39} & \textbf{9} \\
\end{tabular}}
\end{table}

We see significant improvement in ASR over the baseline when implementing RTST in all three adversarial benchmarks, particularly in \textbf{JBB PAIR} and \textbf{JBC + Reddit}, where the baseline model struggles to identify adversarial prompts. RTST improves ASR performance by 60.0\% and 35.0\%, respectively, on these benchmarks, significantly improving the security of the Gemini 2.5 Flash model. Unfortunately, there is also a significant level of framework-caused benign refusals, pointing to an important direction of further research.

\subsection{Ablation Testing}
Next, we conduct ablation testing to examine the performance contributions of each framework component. This includes the use of real-time weight optimization compared to static defaults and the performance of a model optimized over a set of prompts compared to one with default weights.

We implement 4 testing configurations as described in \textbf{Table 5}.

\begin{table}[h!]
\caption{Ablation Testing Configurations}
\centering
\resizebox{\columnwidth}{!}{%
\begin{tabular}{l|p{5.5cm}}
\textbf{Configuration} & \textbf{Description} \\\hline
INIT & RTST with initial Behavior weights and set, real-time optimization deactivated.\\
INIT$_{opt}$ & RTST with initial Behavior weights and set, real-time optimization activated.\\
TRAINED & RTST with Behavior weights and set after weight optimization from \textbf{General Benchmarking}, real-time optimization deactivated. \\
TRAINED$_{opt}$ & RTST with Behavior weights and set after weight optimization from \textbf{General Benchmarking}, real-time optimization activated. \\
\end{tabular}}
\end{table}

For each test, we utilize 400 prompts from the Qualifire Benchmark dataset \cite{qualifire}, with a 50-50 split between benign and adversarial prompts. The subset of prompts is chosen at random and utilized for all four configuration tests, with prompt order randomized for each test. ASR and RR are calculated solely from RTST safety determinations without considering Main Model responses. The results of these tests are shown in \textbf{Table 6.}

\begin{table}[h!]
\caption{Ablation Testing Results}
\centering
\resizebox{\columnwidth}{!}{%
\begin{tabular}{l|c|c|c}
\textbf{Configuration} & \textbf{ASR} & \textbf{RR} & \textbf{F1}\\\hline
Baseline (Gemini 2.5 Flash) & 70.0\% & 6.0\% & 0.451\\
\hline
INIT & 12.0\% & 15.0\% & 0.867\\
INIT$_{opt}$ & 9.0\% & 15.0\% & 0.883\\
TRAINED & 11.5\% & 14.5\% & 0.872\\
TRAINED$_{opt}$ & 10.5\% & 12.0\% & 0.888\\
\end{tabular}}
\end{table}

Changes in ASR and RR are slight between configurations, though still noticeable. In both the INIT and TRAINED configuration pairs, ASR decreased by 1-3\%, RR decreased by 0-2.5\%, and F1 increased by 0.016 when real-time optimization was activated. Similarly, all performance indicators improved when comparing the TRAINED configurations to their INIT counterparts, except for a 1.5\% increase in ASR between the real-time optimized INIT and TRAINED configurations. ASR and F1 improved significantly between all configurations when compared to the baseline, supporting empiricial results from earlier benchmarking.

These results suggest that implementing real-time optimization functionality can enhance our framework's performance, especially when RTST is allowed to self-tune across a wide range of prompts.

\section{Conclusions}
We introduce RTST, a novel framework that takes inspiration from multi-agent systems for LLM security and improves adaptability, learning overhead, and manual-tuning control. Through benchmarking, we demonstrate that a simplified two-agent system can significantly improve ASR and F1 performance when identifying jailbreak prompts while maintaining high explanability. Ablation testing on our framework indicates that real-time, single-prompt optimization of moderator model performance is possible and usable in defending against adversarial attacks. The promising performance of RTST supports the viability of simple, adaptive approaches to LLM security.

Given the performance and novel architecture of the RTST framework, we identify several points of future research. Due to resource constraints in our experimentation, more intensive benchmarking and ablation testing are important next steps for a broader evaluation of the framework's performance. Further improvement to reduce refusal rates would also be useful, perhaps by altering the initial Behavior set. Testing of different hyperparameter values or schemas (such as non-linear weight tuning or adaptive $K$ Behavior sizes) and moderator models would be interesting to improve the rigor and efficacy of RTST's self-tuning techniques. Finally, application of an RTST-similar framework to a task outside of LLM defense may reveal other practical uses of the framework (perhaps in implementing real-time learning for zero-shot classification tasks).

\section*{Acknowledgements}
We thank the Non-Trivial Foundation and Michael Maslowski for guidance, feedback, and support during the research process. We thank Matthew Pisano \cite{bergeron}, Yue Liu \cite{safetysurvey, flipattack, guardreasoner}, and Zikui Cai \cite{aegisllm} for providing expert feedback and direction during ideation.

\bibliographystyle{unsrt}
\bibliography{sample}
\noindent\hfill\vfill\break
\Large\noindent\textbf{Appendices}
\newline\newline\large{\textbf{A. Code Access}}\newline
\noindent\normalsize Access to source code for the project, including system prompts, is available at the following GitHub repository: \textcolor{blue}{https://github.com/EnspikondPlus/rtst}
\newpage
\onecolumn\noindent\large{\textbf{B. Initial Behavior Set and Weights}}\newline
\noindent\normalsize Behaviors are identified with codes for ease of use, with S (supportive), N (neutral), A (adversarial), and an index number. All Behaviors had an initial weight of 1.00.
\begin{table}[h!]
\caption{Initial Behaviors}
\centering
\begin{tabular}{c|p{13.8cm}}
\textbf{Code} & \textbf{Description} \\\hline
S1 & Focuses on building, educating, informing, or generating positive and meaningful content.\\
S2 & Operates within established moral, legal, and safety guidelines without attempting to bypass them.\\
S3 & Reflects values of compassion, inclusivity, respect, and fairness toward all individuals and groups.\\
S4 & Demonstrates a sincere interest in learning, exploring ideas, or gaining insight without manipulation or rhetorical tactics.\\
S5 & Expresses its goals and intent transparently, avoiding misleading or ambiguous phrasing.\\
S6 & Avoids coercive framing and acknowledges the autonomy and dignity of both the model and others.\\
S7 & Lacks any content that suggests, implies, or enables harmful, exploitative, or unethical behavior.\\
S8 & Encourages curiosity and engagement in ways that are mindful of safety, nuance, and ethical implications.\\
S9 & Aims to work together with the model to solve problems, share knowledge, or improve understanding.\\
S10 & Demonstrates accountability and integrity in the use of the model, supporting its responsible deployment.\\
N1 & Does not express any desire to harm others or the model, directly or indirectly.\\
N2 & Raises light ethical questions or ambiguous implications without clearly promoting misuse or wrongdoing.\\
N3 & Uses straightforward language and structure, making the intent and meaning easy to interpret.\\
N4 & Engages with complex or sensitive topics in a way that appears exploratory rather than exploitative.\\
N5 & Deals with subjects that neither clearly align with ethical values nor contradict them.\\
N6 & Aims to perform a task, get information, or interact with the model in a purely instrumental or procedural way.\\
N7 & Poses questions or makes requests without applying pressure or attempting to control outcomes.\\
N8 & Focuses on abstract, fictional, or hypothetical topics without apparent real-world consequences.\\
N9 & Does not take a firm stance on sensitive issues, leaving room for interpretation or discussion.\\
N10 & Avoids loaded language, cultural bias, or topics likely to provoke controversy or conflict.\\
A1 & Seeks to obtain content that is prohibited, harmful, or otherwise restricted in nature.\\
A2 & Uses indirect setups, 'what if' scenarios, or emotional manipulation to bypass safety constraints.\\
A3 & Reformulates prompts, persists through rejections, or uses subtle rewording to evade filters or ethical checks.\\
A4 & Attempts to corner, mislead, or limit the model's choices to compel harmful or unintended output.\\
A5 & Obscures true motives using vague, rhetorical, or deceptive language to sneak past detection.\\
A6 & Advances narratives, ideologies, or actions that cause societal harm, exclusion, or conflict.\\
A7 & Suggests or normalizes actions that are exploitative, unsafe, biased, or morally dubious.\\
A8 & Employs odd formatting, coding, or invented phrasing to confuse safeguards or mask intent.\\
A9 & Tests limits, imposes unnatural constraints, or probes for vulnerabilities in the model's ethical framework.\\
A10 & Constructs prompts to set up future harm, deception, or evasion of content policies.\\
\end{tabular}
\end{table}

\end{document}